 \documentclass[final,5p,times,twocolumn,authoryear]{elsarticle}

\usepackage{amssymb}
\usepackage{lipsum}
\usepackage{color}
\usepackage{amsmath}

\begin{document}

\begin{frontmatter}

\title{Electron heating in bulk overdense plasma aided by time dependent  external magnetic field}

\author{Rohit Juneja$^{1*}$} 
\ead{onlyforjuneja@gmail.com}
\author{Trishul Dhalia$^{1}$}
\author{Amita Das$^{1*}$} 
\ead{amita@iitd.ac.in}

\address{$^{1}$Department of Physics, Indian Institute of Technology Delhi, Hauz Khas, New Delhi 110016, India \\}

\begin{abstract}
%% Text of abstract
This study investigates the localized electron heating in a bulk overdense plasma. The method relies on using a time dependent magnetic field. An initially high external magnetic field imposed on the overdense plasma target enables the propagation of a laser pulse inside it through the pass bands that occur in the magnetized dispersion relation. The choice of decaying external magnetic field is then tailored appropriately to achieve Electron Cyclotron Resonance (ECR)  with the frequency of the laser electromagnetic field. At the resonance location, the field energy of the laser gets transferred to the electrons. These studies have been carried out with the help of the Particle-In-Cell (PIC) simulation technique on the OSIRIS4.0 platform. A detailed study has been carried out to illustrate the energy gain by electrons for a variety of temporal profiles of the magnetic field, laser intensities, and polarizations.  The experiments in this regime may be within reach in the near future. For instance, the choice of long-wavelength CO$_2$ laser requires a magnetic field of about 10s of kilo Tesla to comfortably elicit a magnetized response from electrons. Recent technological advancements have shown the generation of about 1.4 kilo Tesla of magnetic field.

\end{abstract}


\end{frontmatter}

%\tableofcontents

%% \linenumbers

%% main text
\section{Introduction}
\label{sec:Introduction}
The interaction of a laser with the plasma is an important topic of investigation. It has wide ranging applications such as  Particle acceleration (\cite{tajima1979laser,joshi1984ultrahigh,modena1995electron,macchi2013ion}), inertial confinement fusion (\cite{brueckner1974laser,atzeni2004physics,zohuri2017inertial}), generation of x-ray (\cite{rousse2004production,corde2013femtosecond}) and $\gamma$-ray (\cite{cipiccia2011gamma}) sources. The possibility of laboratory astrophysics has also been illustrated (\cite{remington2000review}). Plasma heating is another important aspect. There are many conventional mechanisms of heating the plasma by lasers which rely on collisional and collisionless resonance processes (\cite{estabrook1978properties,ping2008absorption,  wilks1992absorption, chopineau2019identification, freidberg1972resonant, stix1965radiation, gibbon1992collisionless,kaw1969laser,kaw2017nonlinear,das2020laser,yabuuchi2009evidence}). Furthermore, processes which rely on the generation of space charge sheath fields, such as $\vec{J}\times\vec{B}$ heating (\cite{Kruer1985JBHB}), vacuum heating, etc., are also being widely implemented (\cite{Brunel1987NotsoresonantRA}).
These conventional mechanisms are not operative to heat the particles in a bulk overdense localised region of the plasma. In this work, we propose a scheme that can heat electrons in the bulk overdense plasma region. The scheme relies on the application of a time dependent external magnetic field. The dispersion relation of the magnetized plasma offers several resonances. One such resonance is electron cyclotron resonance (ECR) (\cite{geller2018electron}) in which the laser frequency matches with the electron cyclotron frequency in the applied magnetic field. The orientation of the magnetic field in this case is along the laser propagation direction. 
\begin{figure*}
  \centering
  \includegraphics[scale = 0.24]{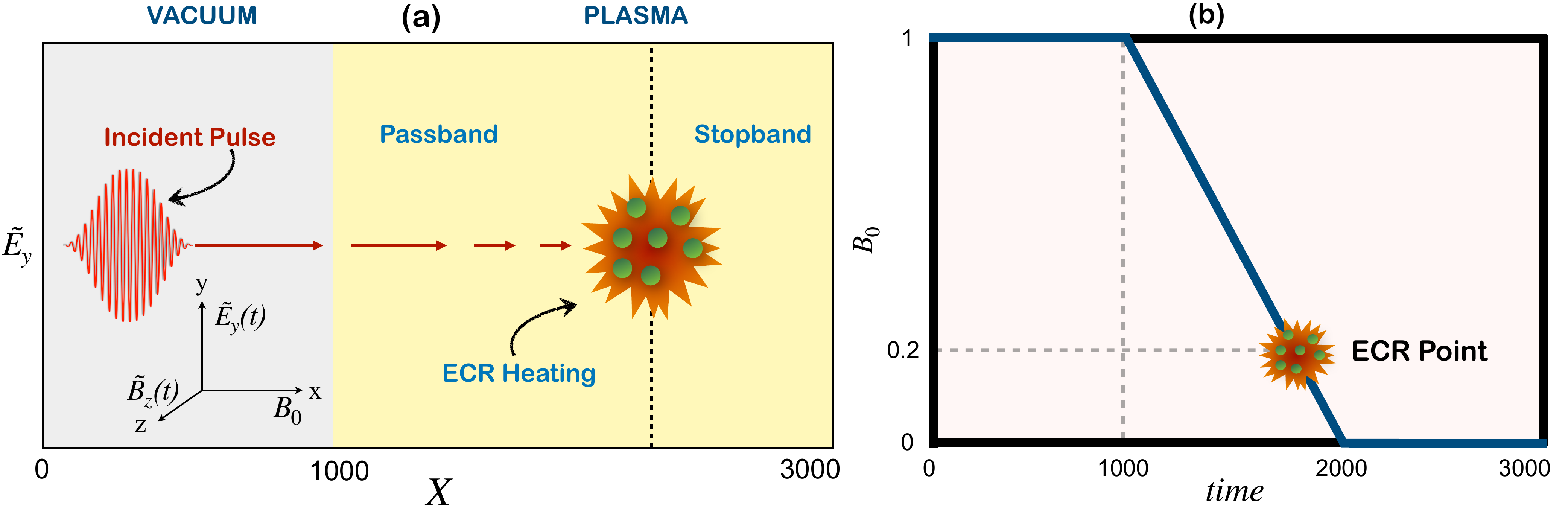}% Images in 100% size
  \caption{Schematic (not to scale) of laser-plasma interaction in the presence of a time-dependent external magnetic field. Subplot (a) shows the laser pulse incident on a magnetized plasma, propagating into the target until it reaches the electron cyclotron resonance (ECR) layer. At this point, the resonance condition $\omega = \omega_{ce}$ is satisfied, which enables efficient energy transfer from the laser to the plasma electrons that results in localized absorption. Subplot (b) illustrates the temporal evolution of the external magnetic field applied. This time dependent magnetic field governs the location and time at which the laser energy is absorbed by the plasma.}
  
\label{Fig:schematic}
\end{figure*}

The Electron Cyclotron Resonance (ECR)  is widely used for heating Tokamak plasmas. Low-frequency (GHz range) EM waves produce ECR resonance for the external magnetic field employed in tokamaks. The geometry of the magnetic field and its spatial variations are quite complex in the case of Tokamaks.  (\cite{bornatici1983electron,erckmann1994electron,laqua1997resonant,ganguli2019evaluation}). A very high value of magnetic field is required to obtain ECR resonance with the frequency of the laser EM field. Currently, magnetic fields of the order of Kilo Tesla have already been produced in laboratory (\cite{nakamura2018record}). An order of magnitude enhancement in the magnetic field will bring the parameters in the right ballpark for observing ECR  and experimentally exploring the mechanism put forth here for the low-frequency CO$_2$ lasers.  It is thus conceivable that with rapid technological advancements (\cite{nakamura2018record,korneev2015gigagauss}), this regime will be experimentally realizable soon. It is for this reason that the regime of magnetized laser plasma interaction has recently drawn considerable research interest (\cite{vashistha2020new, Juneja_2023,maity2022mode,mandal2021electromagnetic,dhalia2023harmonic,goswami2022observations,juneja2024enhanced,dhalia2024absorption}).

We propose to utilize the pass band of the dispersion relation of a magnetized plasma to have the laser EM field propagate inside the overdense plasma medium. 
The magnetic field intensity initially is chosen such that the laser EM field frequency lies in the passband of the dispersion curve of the $R$ mode and stopband of the $L$ mode. This ensures that only the right circularly polarized  EM wave propagates inside the overdense plasma.  As the intensity of the applied external magnetic field decreases with time, the laser frequency then hits the EC resonance layer in the plasma.  This resonance leads to the transfer of field energy to electron kinetic energy.   The temporal profile of the magnetic field can be appropriately tailored to have the ECR resonance at a suitable spatial location for heating. The absorption efficiency for various polarizations of the incident laser pulse, different temporal profiles of the magnetic field, etc., have been studied with the help of 1-D Particle-In-Cell simulations.  Other features, such as the role of EM wave intensity on absorption, are also investigated. 
This paper is structured as follows: Section \ref{sim} provides details of the simulation and the choice of parameters. Section \ref{sec:ResultDiscussions} presents the observations and their analysis. In section \ref{conclusion}, we summarize and conclude our studies.

\begin{table*}
    \caption{Simulation parameters in normalized units and possible values in standard units.} 
  \begin{center}

\setlength{\tabcolsep}{14pt}
  \begin{tabular}{ccc}
  \hline
  \hline
  \hline
      \textbf{Parameters}  & \textbf{Normalized value}   &   {\textbf{Value in standard units}}\\
      \hline
      \hline
      \hline\\
      \hline
    {\textbf{Laser Parameters}} \\
    \hline\\
      Frequency ($\omega_{L}$)  & $0.2 \omega_{pe} $ & $ 0.2 \times 10^{15} Hz$\\
      Wavelength ($\lambda_{L}$)  & 31.4$c/\omega_{pe}$ &9.42 $\mu m$\\
      Intensity ($I_0$)  & $a_0 = 0.05$ & $3.8\times10^{13}Wcm^{-2}$\\\\
      \hline
    {\textbf{Plasma Parameters}} \\
    \hline\\
      Number density ($n_{0}$) & 1 & $3.15 \times 10^{20} cm^{-3}$\\
       Electron plasma frequency ($\omega_{pe}$) & 1 & $10^{15} Hz$\\
       Skin depth ($c/\omega_{pe}$) & $1$ & $0.3{\mu}m$\\\\
       \hline
       {\textbf{Simulation Parameters}} \\
       \hline\\
       $L_x$ & $3000$ & $900 \mu m$\\
       $dx$ & $0.05$ & $15 nm$\\
       $dt$ & $0.02$ & $ 2 \times 10^{-17} s$\\ \\
      \hline
      \hline
      \hline
  \end{tabular}
  \end{center}
  \label{Table}
\end{table*}

\begin{figure*}
  \centering
  \includegraphics[scale = 0.2]{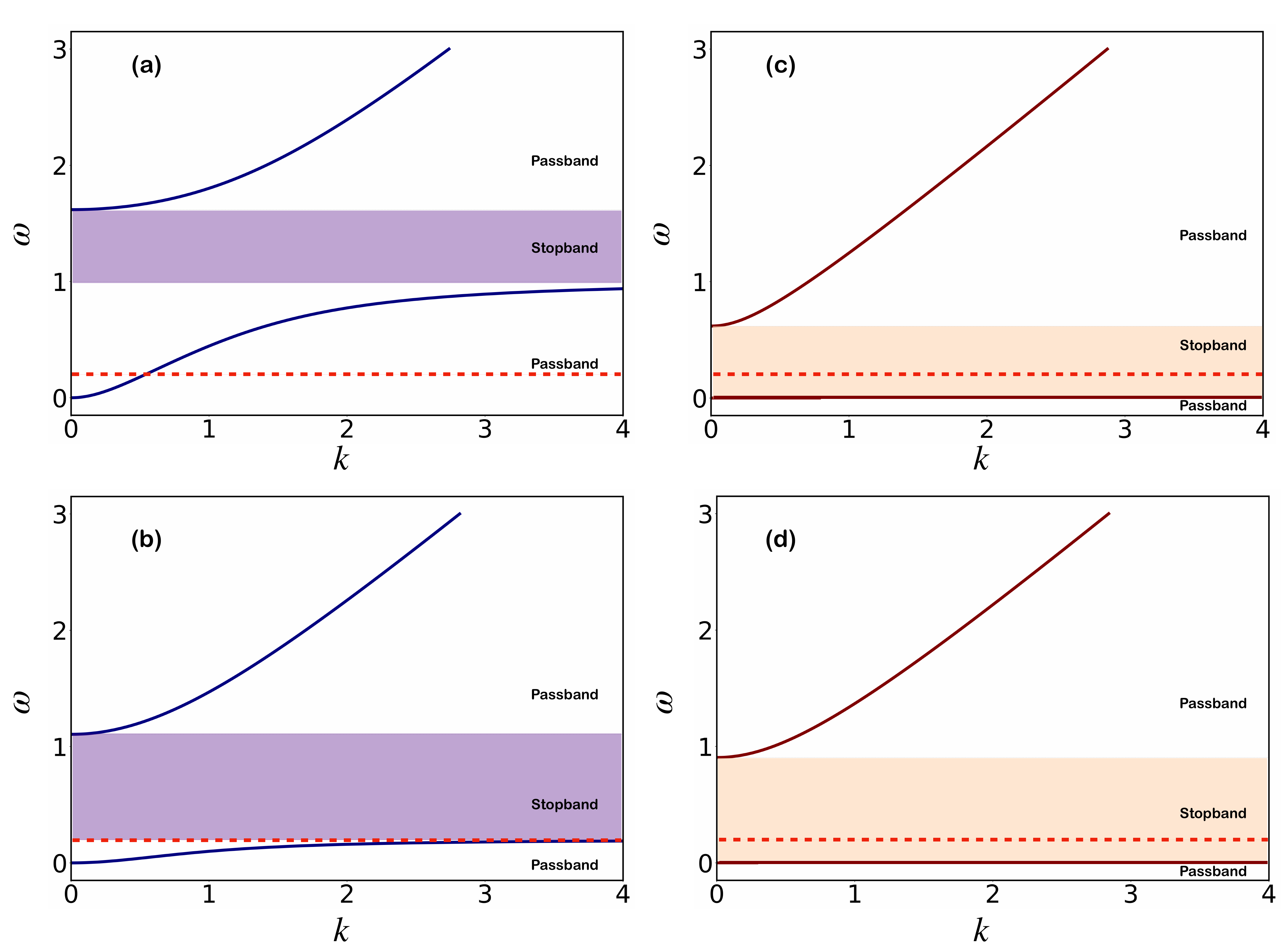}% Images in 100% size
  \caption{Dispersion curves in the RL-mode geometry, showing the passbands and stopbands for different magnetic field strengths. Subplots (a) and (b) show the dispersion curves for the right-hand circularly polarized (R) mode at $B_0 = 1$ and $B_0 = 0.2$, respectively, while subplots (c) and (d) show the corresponding curves for the left-hand circularly polarized (L) mode at the same magnetic field strengths. The plots demonstrate that only the R mode exhibits a passband that allows wave propagation inside the plasma.}
\label{Fig:dispersion_curve}
\end{figure*}
 
\section{Simulation details}
\label{sim}
In this study, one-dimensional (1D) Particle-In-Cell (PIC) simulations have been conducted using a massively parallel PIC code OSIRIS 4.0 (\cite{hemker2000particle, fonseca2002osiris, fonseca2008one}). A one-dimensional simulation box with a length of $L_x = 3000 c/\omega_{pe}$ has been considered. Here, $c$ denotes the speed of light in a vacuum, while $\omega_{pe}$ represents the electron-plasma frequency. In all our simulations the time has been normalized by  $t_{n}$ = ${\omega_{pe}}^{-1}$. For length, the skin depth  $x_{n} = c/\omega_{pe}$ has been chosen. Normalization of the electric and magnetic fields is done by $E_{n} = B_{n} = m_{e}c\omega_{pe}/e$, where $m_{e}$ is the mass of the electron and $e$ represents the magnitude of the electronic charge. Absorbing boundary conditions have been taken for both fields and particles in both directions. We have considered $60000$ grid points (cells) in our simulations, which correspond to the grid size $dx = 0.05c/\omega_{pe}$, while the temporal resolution is chosen to be $dt = 0.02\omega_{pe}^{-1}$.  The number of particles per cell is taken to be 8. From $x=0$ to $1000c/\omega_{pe}$, there is a vacuum region, and the plasma boundary starts from  $1000c/\omega_{pe}$. These have also been shown in Table \ref{Table}, along with the possible realistic values that this choice of parameters can correspond to. The schematic of the simulation geometry is shown in Fig. \ref{Fig:schematic}. 
A laser pulse with an intensity of $3.8 \times 10^{13} W cm^{-2}$ (corresponding to a normalized vector potential of $a_{0}=eE_{l}/m\omega_{l}c=0.05$) was selected for the linearly polarized case. For comparison with circular polarization studies, a value of $a_{0}=0.05/\sqrt{2}$ was chosen to ensure that the total energy of the incident laser is identical in both linearly and circularly polarized cases.
The laser pulse propagation is in the $\hat{x}$ direction. The electric and magnetic field components of the laser pulse are in $\hat{y}$ and $\hat{z}$ direction, respectively. The applied magnetic field is also along the $\hat{x}$ direction. It is static till $t = 1000$, and falls linearly with time as depicted in Fig. \ref{Fig:schematic}. This geometry essentially supports the RL modes. Even though the plasma is overdense, the presence of the externally applied magnetic field permits the propagation of electromagnetic waves when their frequency lies in the pass band of the dispersion curve depicted in Fig. \ref{Fig:dispersion_curve}. The (a) and (b) subplots have been drawn for the magnetic field values $B_0 = 1$ and $B_0 = 0.2$, respectively, for the R-mode dispersion. The (c) and (d) subplots have been drawn for the magnetic field values $B_0 = 1$ and $B_0 = 0.2$, respectively, for the L-mode dispersion. The laser frequency lies in the passband of the R-mode and stopband of the L-mode dispersion curves for $B_0 = 1$. For $B_0 = 0.2$, the laser frequency hits the electron cyclotron resonance for R-mode dispersion, and it lies in the stopband of L-mode dispersion. Thus, the external magnetic field allows the laser to penetrate inside the bulk of the overdense plasma target.

\begin{figure}
  \centering
  \includegraphics[scale = 0.13]{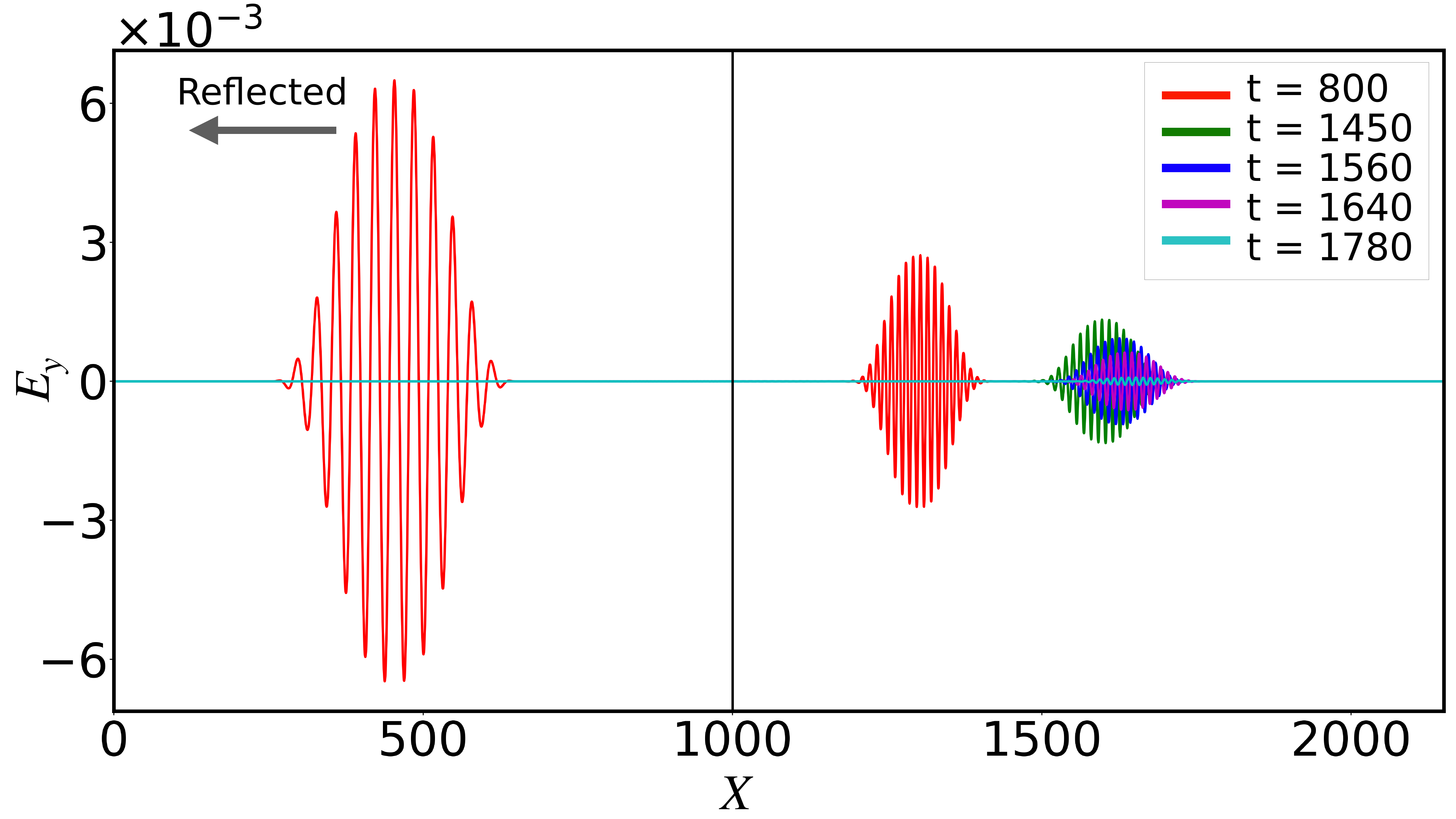}% Images in 100% size
  \caption{The $y$-component of the electric field ($E_y$) plotted as a function of $x$ at various time steps during the simulation. The different curves, indicated by various colors, represent snapshots of the electric field at different simulation times ranging from $t = 800$ to $t = 1780$. This temporal evolution shows how the field structure evolves as the laser propagates through the plasma.}
  % \caption{The $y$ component of the electric field ($E_y$) versus $x$ has been shown at different times of the simulation runs. Here, the different colours show the $x$ locations of the fields at different times of the simulation run ($t$ = $800$ to $1780$).}
\label{Fig:EyEvolution}
\end{figure}

% \textcolor{red}{edited upto here}
\section{Observations} \label{sec:ResultDiscussions}
This section presents the simulation results along with their interpretation. The incident laser pulse is plane-polarized, with its electric field oriented along the $\hat{y}$ direction. The plasma is overdense for the choice of the laser frequency, however, this frequency falls within the passband of the R-mode dispersion curve, allowing the wave to propagate within the plasma.

Fig. \ref{Fig:EyEvolution} shows a snapshot of $E_y$ at various times. As the pulse encounters a time-dependent magnetic field inside the plasma, the reduction in the amplitude of the $E_y$ component is observed. When the laser pulse hits the plasma surface, one part of the pulse gets reflected, and the other enters the plasma. The figure shows these components at $t=800$, after they have propagated some distance from the interaction point.

\begin{figure}
  \centering
  \includegraphics[scale = 0.16]{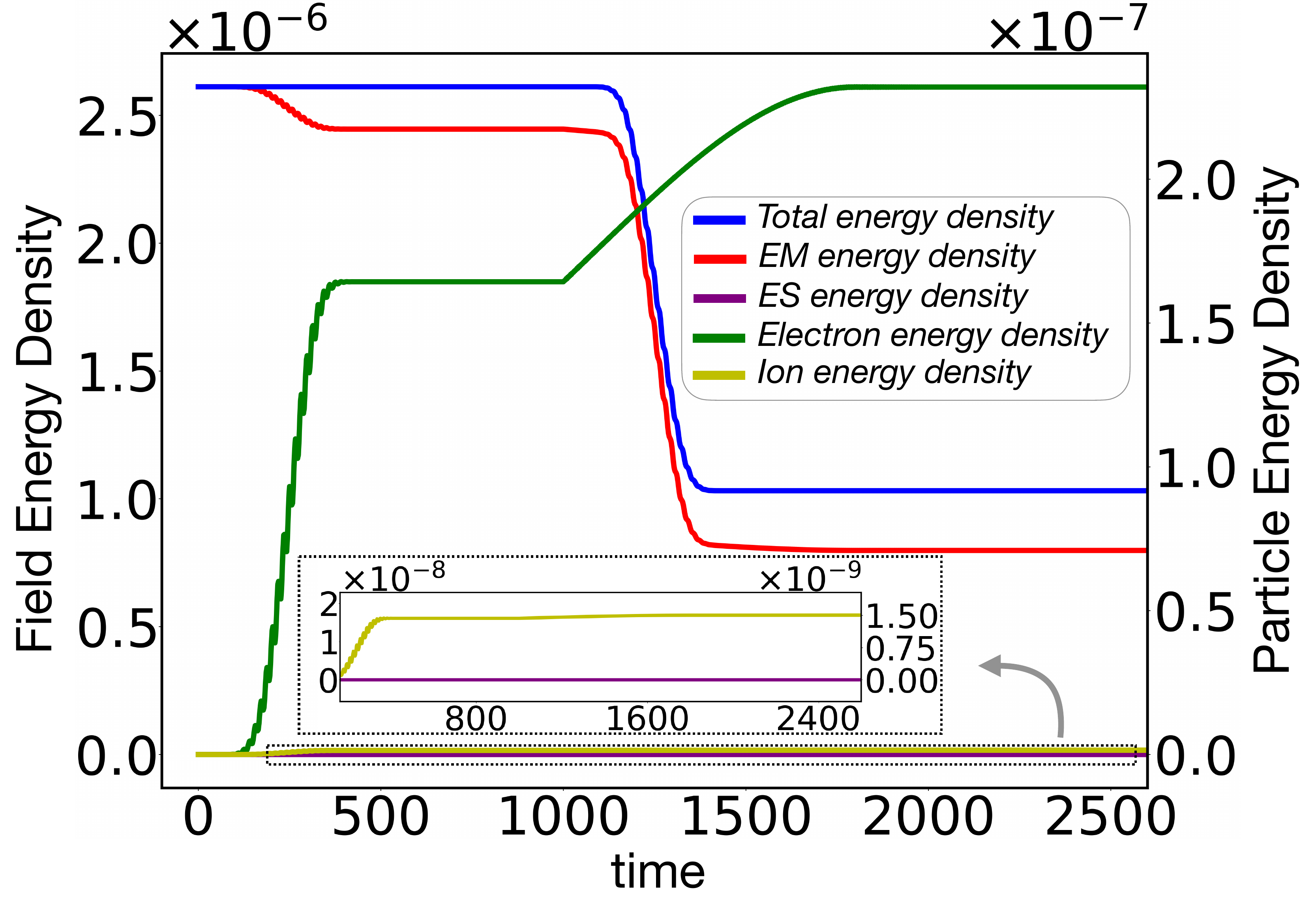}% Images in 100% size
  \caption{Time evolution of the spatially averaged energy densities for different components of the system. The plot shows the variation of electron energy, ion energy, electromagnetic field energy, electrostatic field energy, and the total energy density as a function of time. This provides understanding into the dynamics of energy transfer between fields and particles during the laser-plasma interaction, as well as the overall energy conservation throughout the simulation.}
  % \caption{Time evolution of spatially averaged Electron, Ion, Electromagnetic, Electrostatic, and total energy density.}
\label{Fig:EnergyEvolution}
\end{figure}

In  Fig. \ref{Fig:EnergyEvolution}, the evolution of various energies over time has been shown. Time is represented on the $x$-axis, while the left and right $y$-axis represent the field energy density and particle energy density, respectively. The total energy, shown by the blue solid line, remains conserved throughout the simulation. There is, however, a dip at $ t = 1150$, which occurs when the reflected pulse from the plasma vacuum boundary moves out of the left boundary. 
Initially, all the energy resides in the electromagnetic fields and is shown by the red solid line. As time progresses, this electromagnetic energy gradually decreases, while the kinetic energy of the electrons (green solid line) increases correspondingly. During this period, only the electrons exhibit significant energy gain, and the ions acquire negligible energy.  At time $t = 1640$, the laser pulse finally reaches the electron cyclotron resonance (ECR) layer, and the EM wave propagation stops completely, with group velocity being zero at the resonance layer. There is a direct transfer of this electromagnetic energy to the kinetic energy of electrons without the generation of electrostatic energy. This is discussed in the paper \cite{Juneja_2023}, where in RL mode configuration, there is no generation of electrostatic energy and particles gain energy directly from the electromagnetic resonance (here, ECR in this case). The purple solid line represents the electrostatic energy, which has a zero value throughout the evolution. In the region around the resonance, where this electromagnetic energy is dumped into the kinetic energy of electrons, localized heating of plasma takes place.

\begin{figure}
  \centering
  \includegraphics[scale = 0.14]{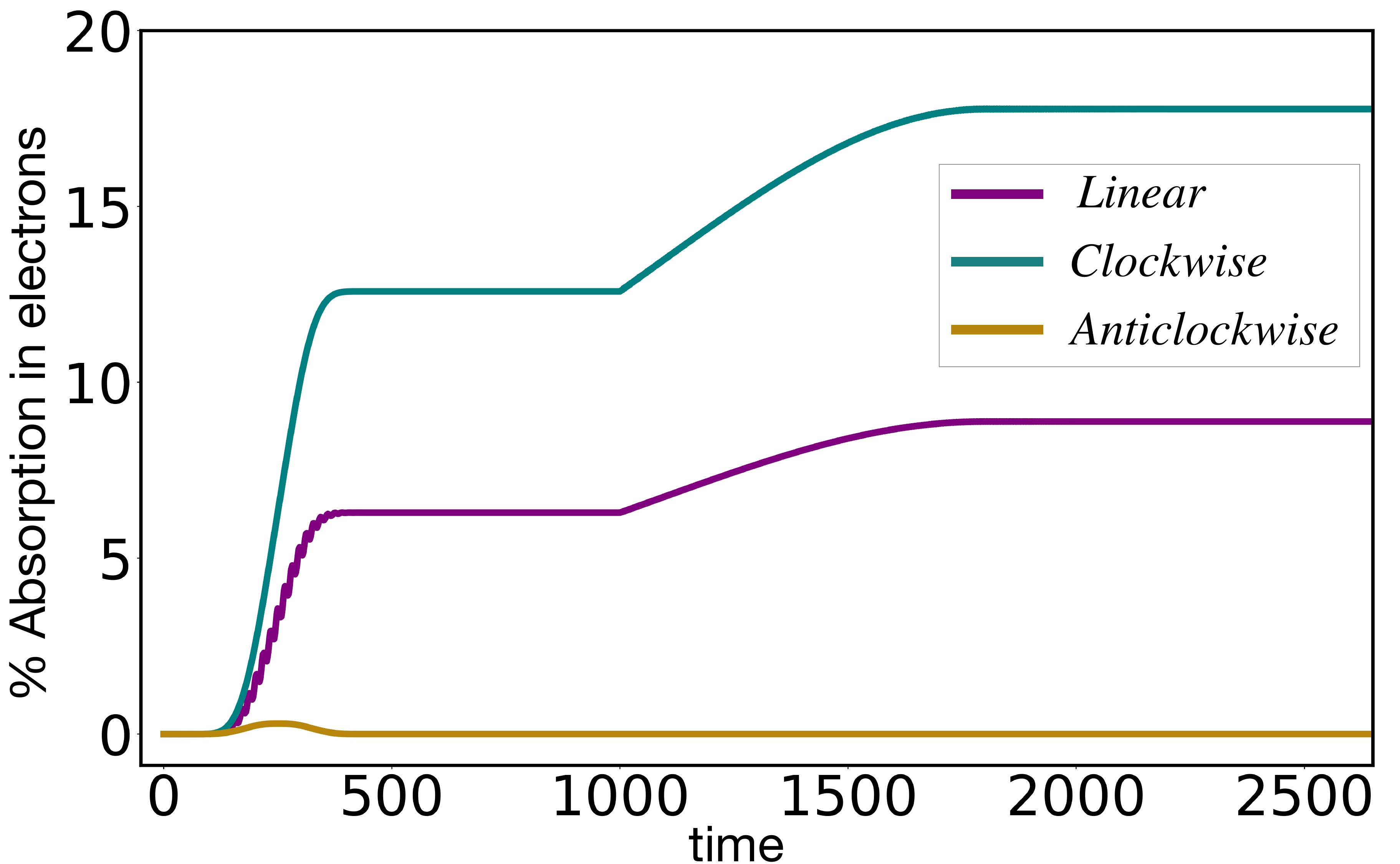}% Images in 100% size
  \caption{Time evolution of the spatially averaged percentage absorption in electron energy for different laser pulse polarizations. The plot compares the absorption efficiency among linearly polarized, clockwise circularly polarized, and anticlockwise circularly polarized laser pulses. This comparison underlines the dependence of energy transfer on the laser polarization in the presence of an external magnetic field and shows how different polarization states affect the coupling between the laser and the plasma electrons over time.}
  % \caption{Time evolution of spatially averaged \% Absorption Electron Energy comparison among Linear, Clockwise, and Anticlockwise laser pulse.}
\label{Fig:PolarizEnergies}
\end{figure}

\begin{figure}
  \centering
  \includegraphics[scale = 0.2]{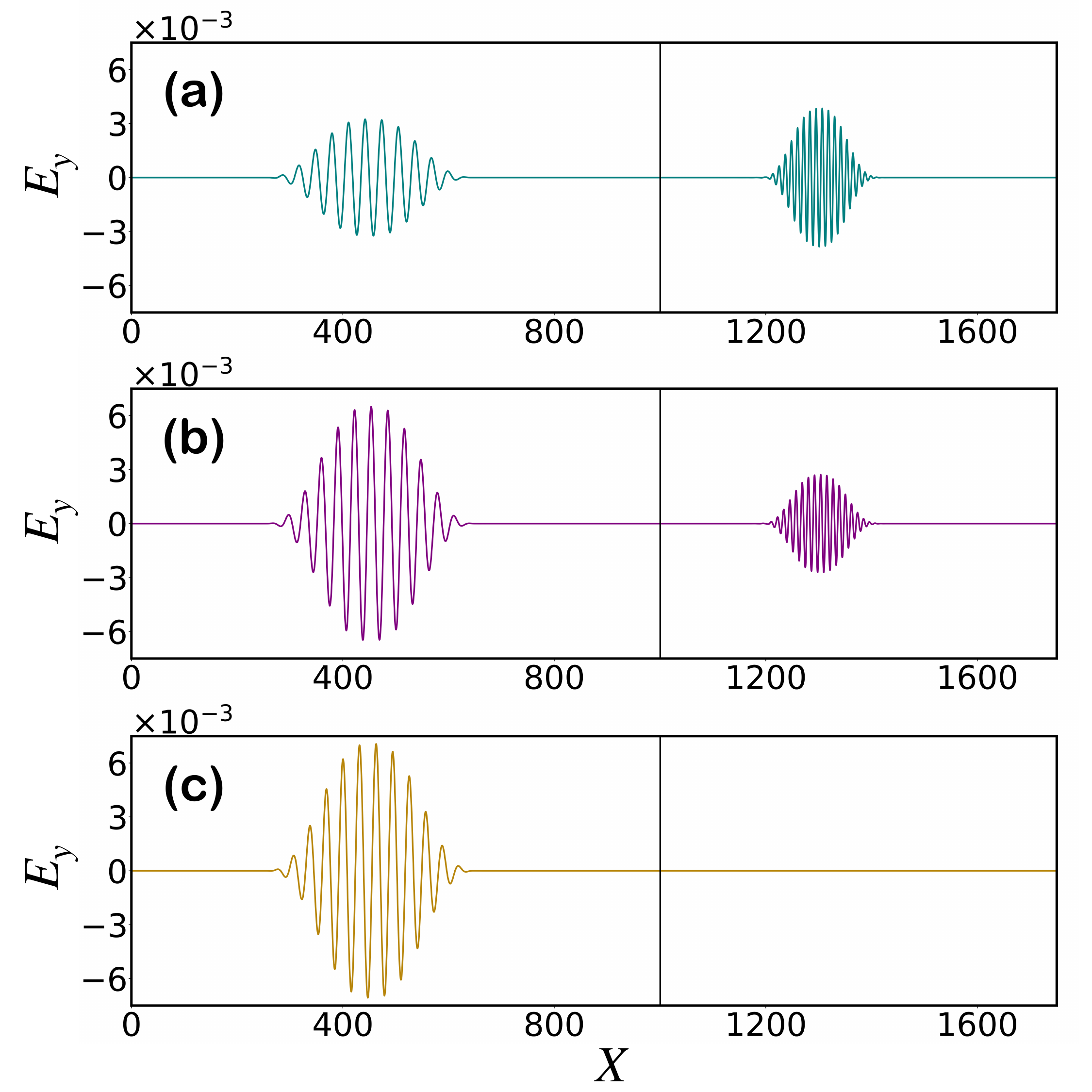}% Images in 100% size
  \caption{The $y$-component of the electric field ($E_y$) as a function of $x$, shown at time $t = 800$ for three different laser polarizations. Subplots (a), (b), and (c) correspond to clockwise circular, linear, and anticlockwise circular polarizations, respectively.}
  % \caption{The $y$ component of the electric field ($E_y$) versus $x$ has been shown at $t = 800$ for different polarizations (a) clockwise, (b) linear, and (c) anticlockwise.}
\label{Fig:EyPolariz}
\end{figure}

Extending this analysis, we explore how the polarization of the incident laser pulse influences electron energy absorption. To investigate this, we consider three types of laser polarizations, I) Linear, II) Clockwise (right-hand circularly polarized), and III) Anticlockwise (left-hand circularly polarized).
In Fig. \ref{Fig:PolarizEnergies}, the evolution of the kinetic energy of electrons over time has been shown for three cases, (i) when the laser pulse is linearly polarized, (ii) when it is clockwise polarized, and (iii) when it is anticlockwise polarized. Clearly, the absorption in electrons is highest when the laser pulse is clockwise polarized (right-hand circularly polarized), i.e., when the electric field rotates in the same direction as the electron gyro-motion. For the linearly polarized case, \% absorption lies between the other two cases, because it effectively represents a superposition of left- and right-hand circularly polarized components.
Fig. \ref{Fig:EyPolariz} further supports this observation by showing the amplitude of the electromagnetic field component (specifically, $E_y$) within the plasma for each polarization. For the right-hand circularly polarized (clockwise) case, the amplitude of the $E_y$ component in the plasma is highest, followed by the linear case, and there is no component present in the case of the left-hand circularly polarized pulse. Since electrons gain energy directly from the electromagnetic component, the absorption is maximum for the case when the laser pulse is right-hand circularly polarized.

\begin{figure}
  \centering
  \includegraphics[scale = 0.12]{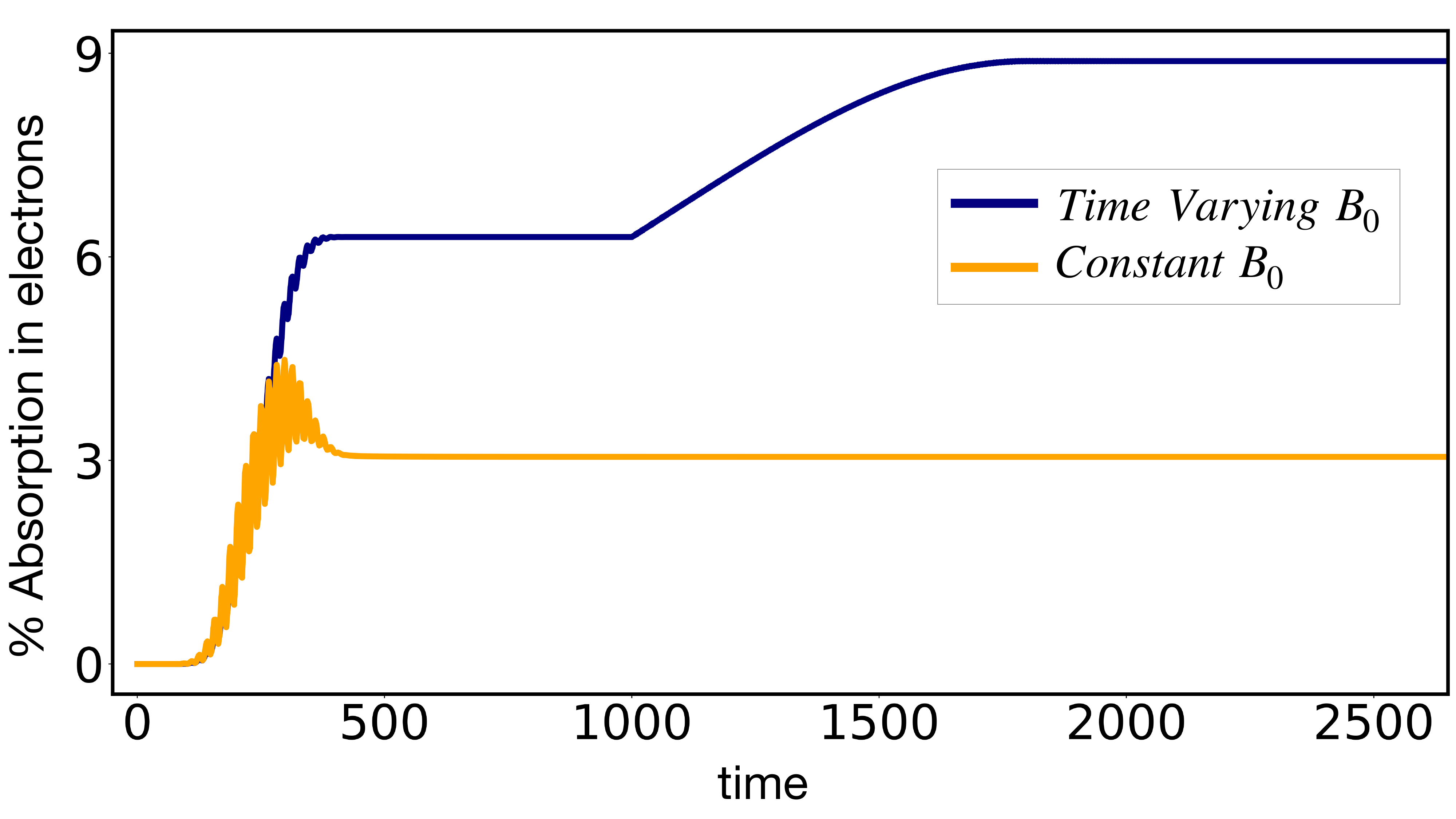}% Images in 100% size
  \caption{Time evolution of the spatially averaged percentage absorption in electron energy, comparing the case of a time-dependent external magnetic field with a case where the magnetic field is held constant at $B_0 = 0.2$. This comparison highlights the role of magnetic field variation in enhancing electron energy absorption.}
  % \caption{Time evolution of spatially averaged \% Absorption Electron Energy comparison with the case where the external magnetic field is constant with $B_0~=~0.2$.}
\label{Fig:EnergyCompareB0}
\end{figure}

\begin{figure}
  \centering
  \includegraphics[scale = 0.14]{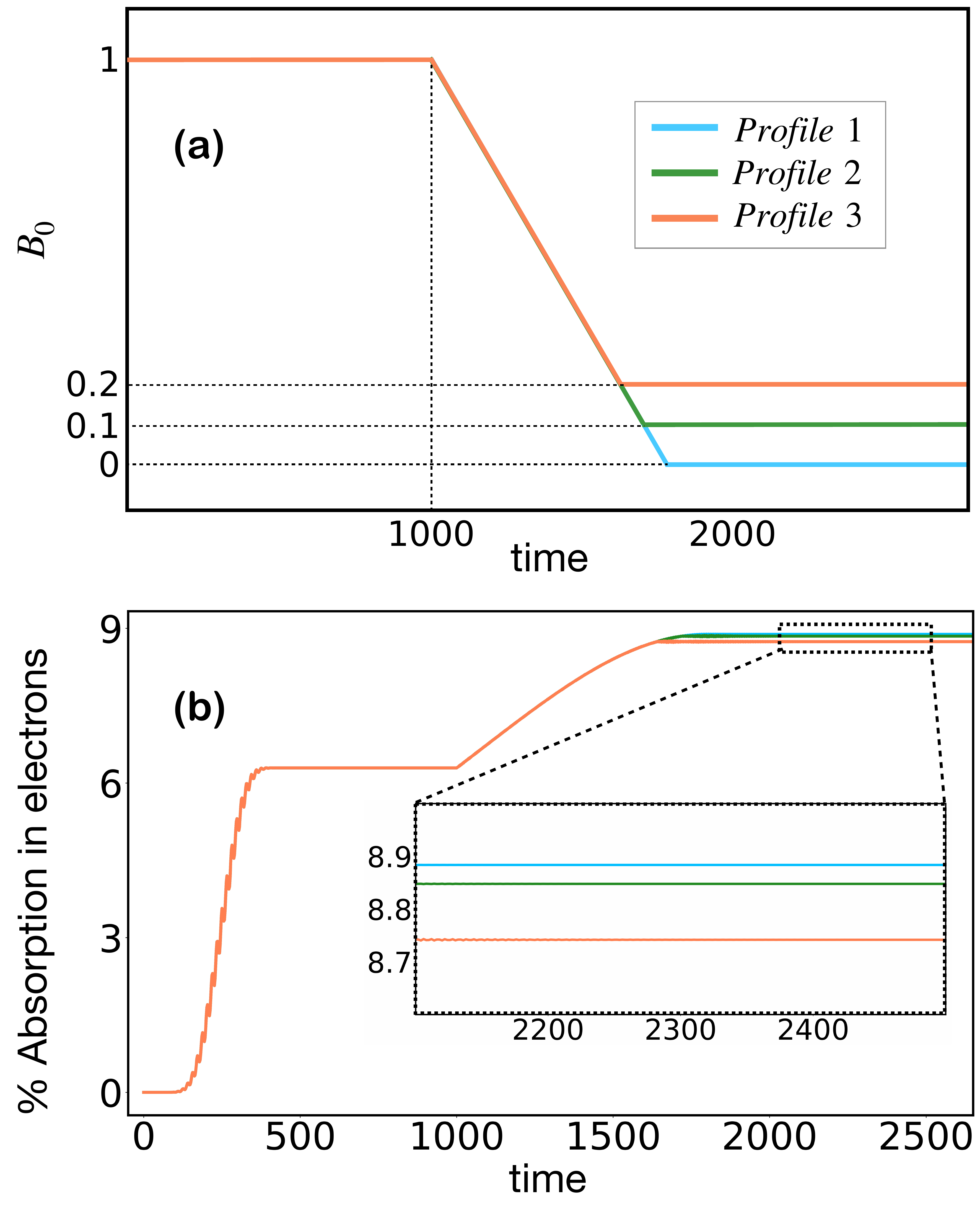}% Images in 100% size
  \caption{Subplot (a) shows different temporal profiles of the external magnetic field applied during the simulation, each profile has the same slope but different saturation levels. Subplot (b) presents the corresponding time evolution of the spatially averaged electron energy density for each magnetic field profile. This comparison shows how variations in the magnetic field’s saturation level influence the energy absorption.}
  % \caption{Various magnetic field profiles shown in (a) and the time evolution of electron energy density with these profiles shown in (b).}
\label{Fig:profilesupto0}
\end{figure}

Following this, we study the effect of an external static magnetic field on energy absorption as shown in Fig. \ref{Fig:EnergyCompareB0}. Here, the time evolution of percentage absorption in electrons is present for the two scenarios, (i) a time-varying external magnetic field (as shown earlier in the Fig. \ref{Fig:schematic}), and (ii) a constant/static external magnetic field of value $B_0=0.2$, which corresponds exactly to the magnetic field required for electron cyclotron resonance. In the first case, the laser pulse initially encounters a stronger value of the magnetic field and gradually reaches towards the resonance point. Whereas, in the second case, the laser interacts with a resonant condition from the very beginning. Clearly, Fig. \ref{Fig:EnergyCompareB0} shows that when a time-varying external magnetic field is used, the energy absorption is around $9\%$, which is significantly greater than $3\%$ in the case when a constant magnetic field is used. This shows that using a time-varying magnetic field can lead to enhanced energy absorption.

\begin{figure}
  \centering
  \includegraphics[scale = 0.14]{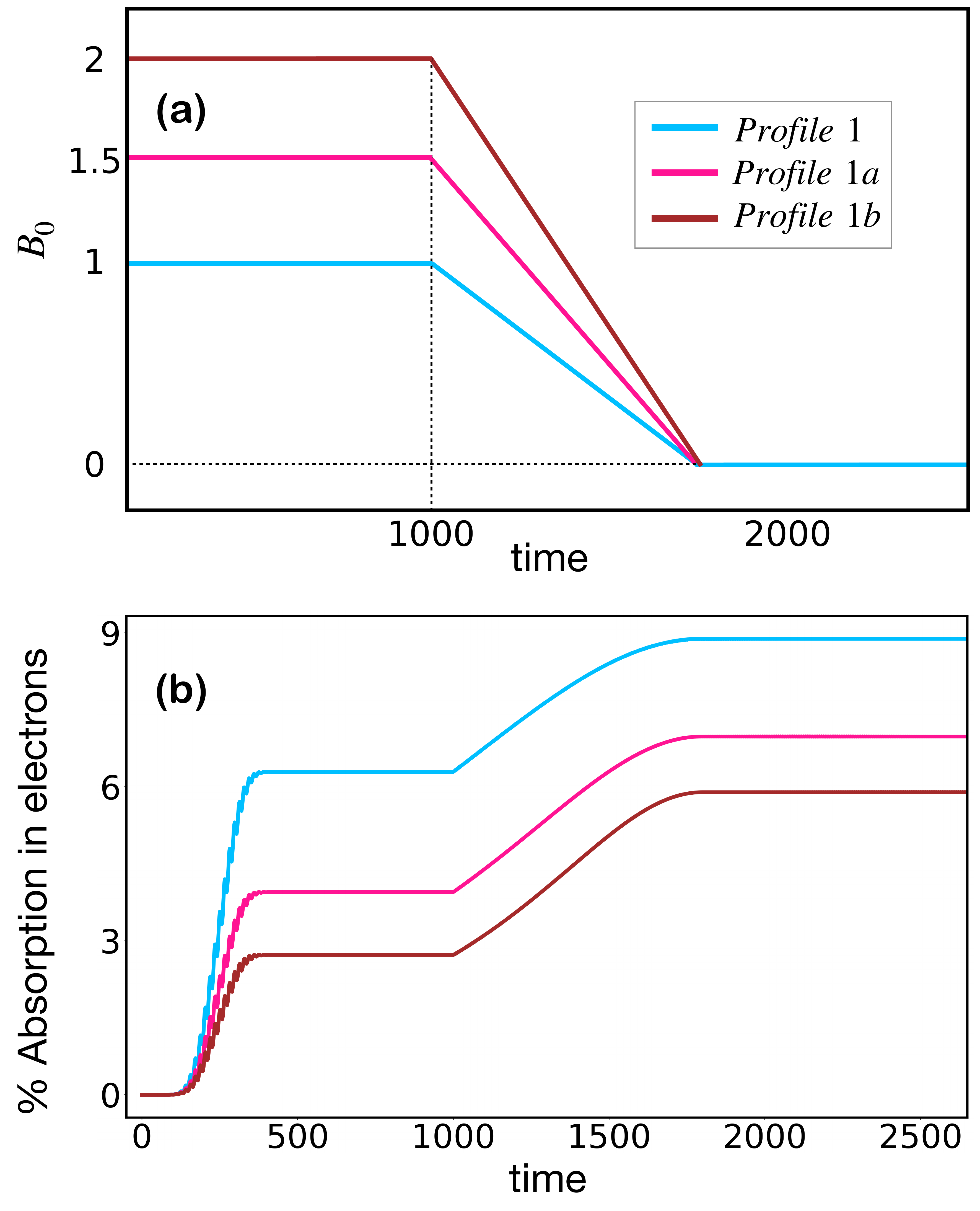}% Images in 100% size
  \caption{Subplot (a) shows different temporal profiles of the external magnetic field applied during the simulation, each profile has a different slope, with the difference in total magnetic field change. Subplot (b) presents the corresponding time evolution of the spatially averaged electron energy density for each magnetic field profile. This comparison shows how variations in the total magnetic field change lead to differences in energy absorption.}
  % \caption{Various magnetic field profiles shown in (a) and the time evolution of electron energy density with these profiles shown in (b).}
\label{Fig:ProfileswithMaxB0}
\end{figure}

Since a time-varying magnetic field leads to better absorption, we now study the effect of different external magnetic field profiles on the absorption process. Fig. \ref{Fig:profilesupto0}(a) illustrates three distinct temporal profiles of the external magnetic field, each evolving differently with time. Profile 1 decays all the way to $B_0 = 0$, while Profile 2 and Profile 3 saturate at values of $B_0 = 0.1$ and $B_0 = 0.2$, respectively, the latter coinciding with the laser frequency, which indicates an electron cyclotron resonance (ECR) configuration. Corresponding to these magnetic field profiles, Fig. \ref{Fig:profilesupto0}(b) shows the percentage of laser energy absorbed by electrons over time. Initially, at $t = 0$, absorption is zero as the laser is yet to interact with the plasma. As time progresses and the laser enters the plasma under a constant magnetic field ($B_0 = 1$), electrons absorb energy, leading to a steady increase in absorption that saturates by the end of this interaction phase. After $t = 1000$, as the external magnetic field begins to decrease based on the defined profiles, we observe a further rise in absorption. The extent of this additional absorption is directly linked to the change in the magnetic field, as the larger the drop in $B_0$ (i.e., greater $\Delta B$), the higher the percentage of energy absorbed by the electrons. This trend is clearly reflected across all three profiles.

\begin{figure}
  \centering
  \includegraphics[scale = 0.14]{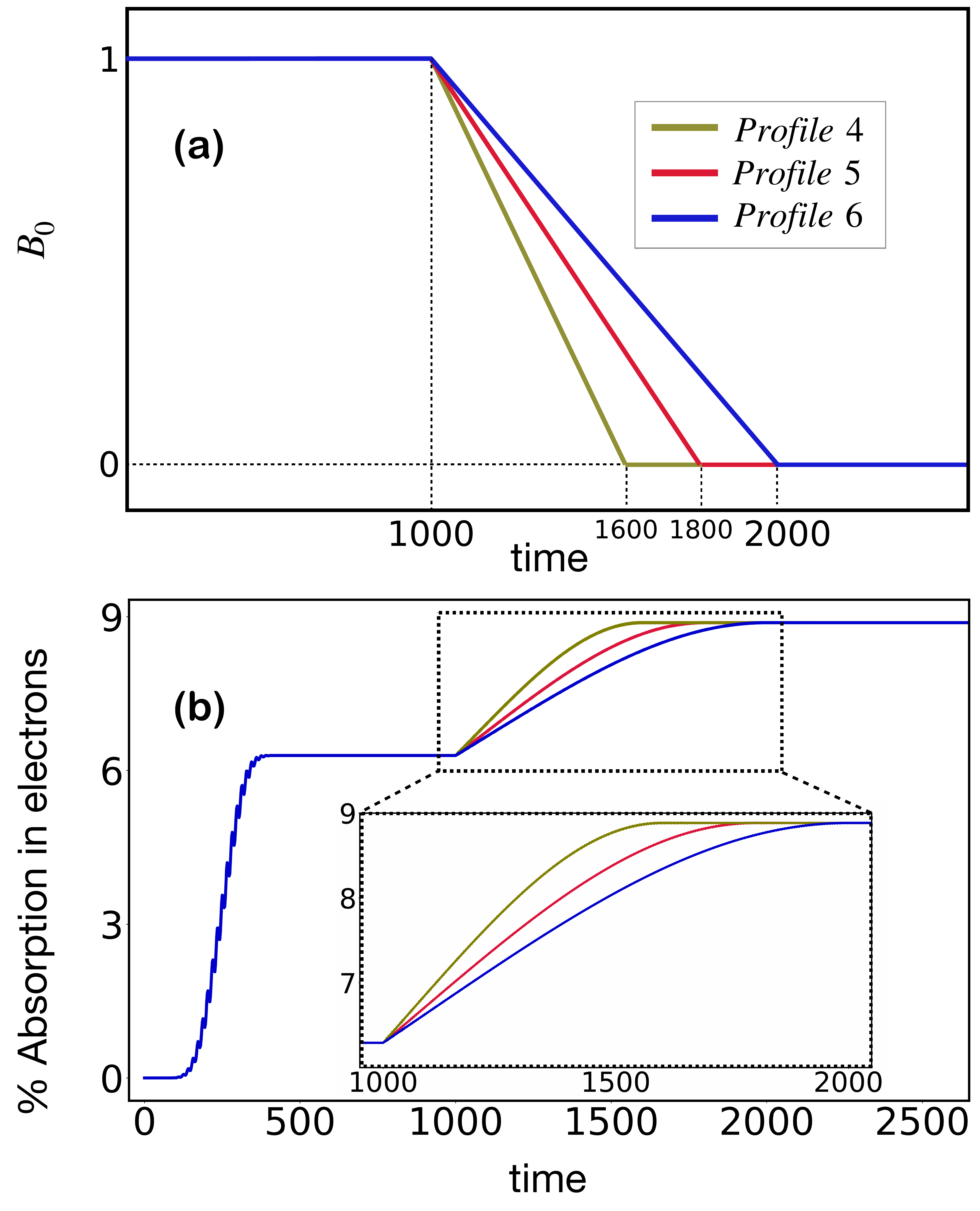}% Images in 100% size
  \caption{Subplot (a) shows a set of magnetic field profiles with different slopes. Subplot (b) shows the corresponding time evolution of the spatially averaged electron energy density for each profile. The comparison shows how the rate of change of external magnetic fields affects the absorption paths.}
  % \caption{Various magnetic field profiles shown in (a) and the time evolution of electron energy density with these profiles shown in (b).}
\label{Fig:profilesSlopes}
\end{figure}

Fig. \ref{Fig:ProfileswithMaxB0}(a) presents three magnetic field profiles that all eventually decay to $B_0 = 0$, but differ in their initial values. The blue curve, Profile $1$, is the same as that shown earlier, beginning from $B_0 = 1$. In contrast, Profile 1a and Profile 1b start from higher initial fields of $B_0 = 1.5$ and $B_0 = 2$, respectively. Fig. \ref{Fig:ProfileswithMaxB0}(b) shows the corresponding laser energy absorption by electrons under these evolving magnetic fields. At $t = 0$, as expected, energy absorption is zero. As the laser interacts with the plasma under different constant values of magnetic field, electrons absorb energy in accordance with how close the initial field is to the ECR condition ($\omega = \omega_{ce}$, which corresponds to $B_0 = 0.2$). Since Profile 1 starts closest to this resonance, it shows the highest energy absorption during the interaction phase, followed by Profiles $1a$ and $1b$. After $t = 1000$, as the magnetic field decreases, all profiles exhibit further energy gain but along different slopes. Interestingly, although Profile $1$ still ends with the highest overall absorption due to its proximity to resonance at the start, the energy gained specifically during the decreasing phase of the magnetic field ($\Delta B$) is largest for Profile $1b$ ($3.2\%$), followed by $1a$ ($3.0\%$), and then $1$ ($2.6\%$). This supports the consistent observation made in Fig. \ref{Fig:profilesupto0}, that a larger change in magnetic field leads to enhanced energy absorption during the decay phase.

Fig. \ref{Fig:profilesSlopes}(a) shows three external magnetic field profiles, all starting from $B_0 = 1$ and decaying down to $B_0 = 0$, but over different durations. This setup ensures that the total change in magnetic field ($\Delta B$) remains the same across all cases. As a result, the final energy absorption by electrons, shown in Fig. \ref{Fig:profilesSlopes}(b), converges to the same value for all three profiles. However, due to the differing rates of magnetic field decay, the evolution of energy absorption follows distinct temporal paths. The profile with the steepest decline in $B_0$ causes the system to reach its saturation value of energy absorption more rapidly, while the more gradual decays result in a delayed approach to the same final state. This clearly demonstrates that while the extent of magnetic field change governs the total energy absorbed, the rate of change controls how quickly that energy is deposited into the plasma.

\begin{figure}
  \centering
  \includegraphics[scale = 0.26]{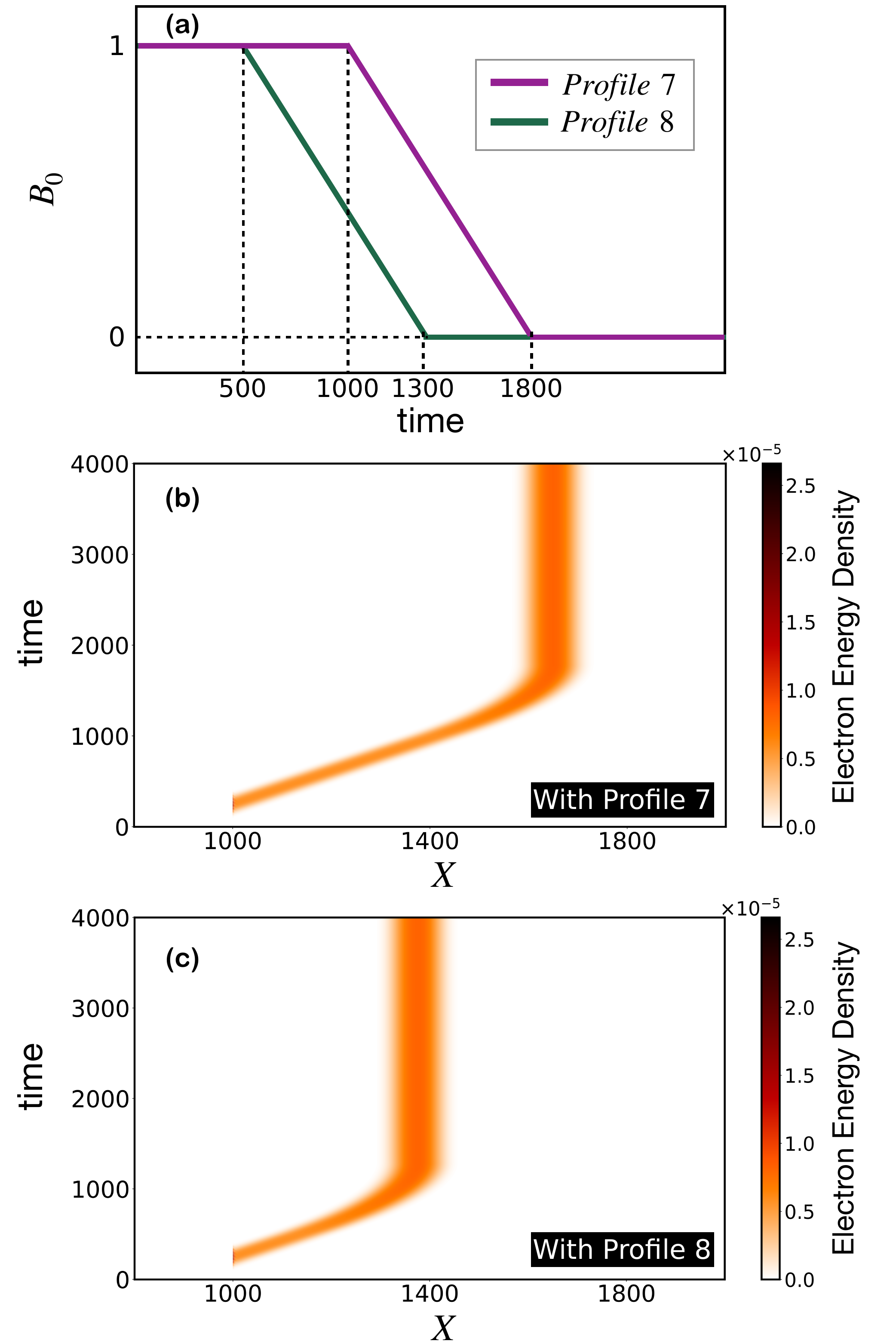}% Images in 100% size
  \caption{Localized electron heating using tailored magnetic field profiles. Subplot (a) shows two different magnetic field profiles that are designed to control the location of resonance within the plasma. Subplots (b) and (c) show the resulting electron energy density, representing localized heating at specific regions.}
  % \caption{Heating at desired location by tailored magnetic field profiles in (a) and corresponding localized heating in (b), and (c).}
\label{Fig:DesiredHeat}
\end{figure}

Based on this, we now discuss how the location of heating within the plasma can be controlled by tailoring the external magnetic field. Fig. \ref{Fig:DesiredHeat}(a) shows two different profiles of the external magnetic field, profile $7$ remains static at $B_0=1$ until $t=1000$, then decreases linearly to $B_0=0$ at $t=1800$ and remains $0$ afterwards. In contrast, the profile $8$ remains static at $B_0=1$ until $t=500$, then linearly falls to $B_0=0$ at $t=1300$, and remains $0$ thereafter. It is interesting to note that the amount and rate of decrease are the same in both profiles, resulting in the same amount of absorption for both profiles. However, as seen in subplots (b) and (c) of Fig. \ref{Fig:DesiredHeat}, the energy absorption is localized in the region around $x=1700$ for profile $7$, and around $x=1400$ for profile $8$. This shows that although the amount of absorption by electrons is the same, different spots in the plasma can be heated by adjusting the profile of the external magnetic field.

\begin{figure}
  \centering
  \includegraphics[scale = 0.2]{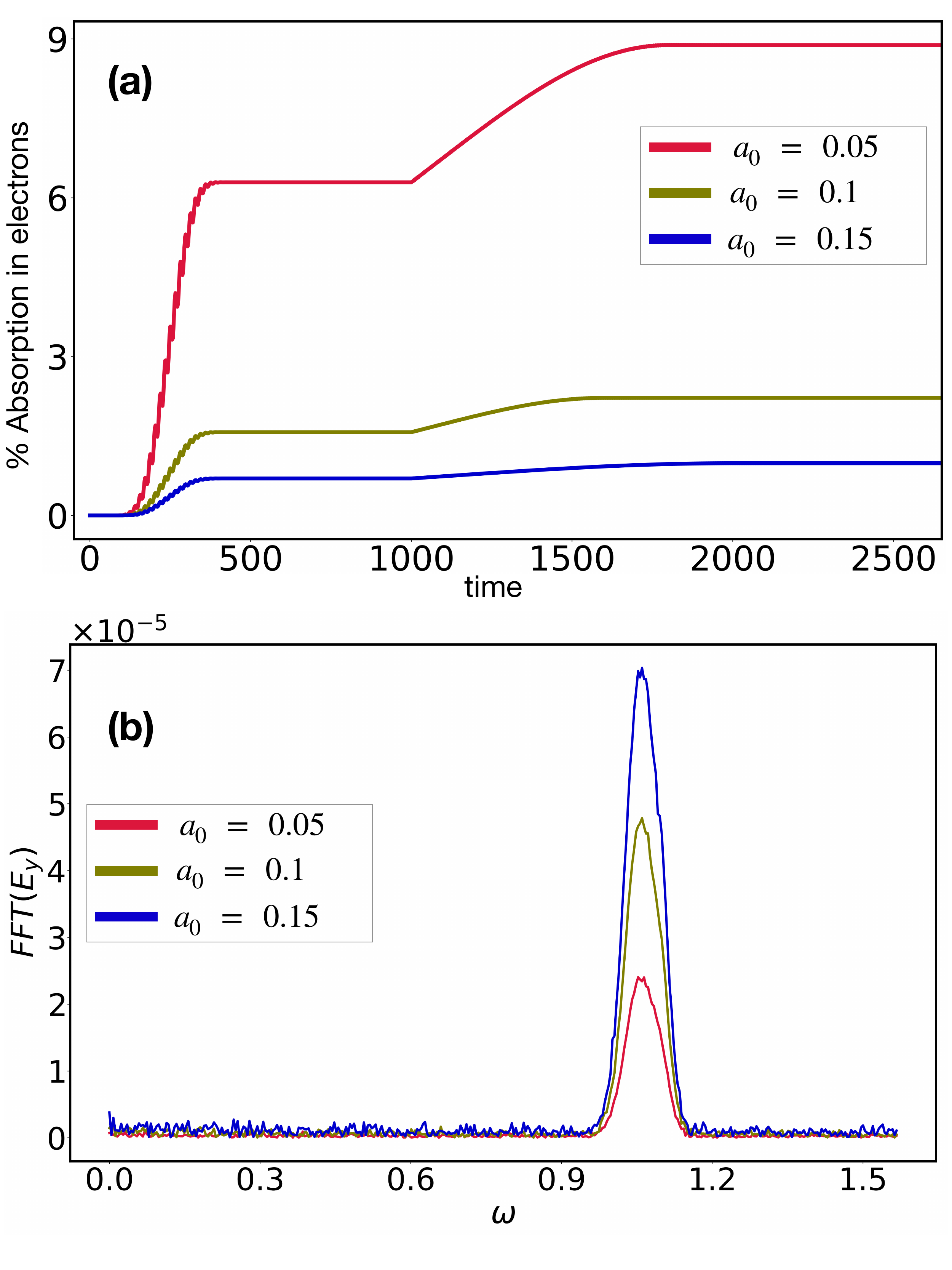}% Images in 100% size
  \caption{Subplot (a) shows the time evolution of spatially averaged percentage absorption in electron energy for different laser pulse intensities. Subplot (b) represents the leakage of higher harmonics from the plasma at various intensities. The results indicate that more intense laser pulses tend to generate stronger harmonic emission and hence carry away more energy, leading to reduced energy absorption into the electrons.}
\label{Fig:IntensityAbsorption}
\end{figure}

After exploring the effects of various magnetic field profiles, we now examine the effect of laser pulse intensity on electron energy absorption. In Fig. \ref{Fig:IntensityAbsorption}, subplot (a) shows the temporal evolution of spatially averaged percentage absorption for three intensities: $a_0 = 0.05$, $0.1$, and $0.15$, while subplot (b) presents the corresponding higher harmonic emission from the plasma. It is observed that as the laser intensity increases, the percentage absorption in electrons decreases. This behavior is linked to the enhanced generation of higher harmonics at higher intensities, which escape from the plasma and carry away a portion of the incident energy. As a result, less energy remains available for resonant absorption into the electrons, leading to reduced absorption at higher intensities.

\section{Conclusion\label{conclusion}}
This work discusses the localized electron heating in a magnetized plasma using electron cyclotron resonance (ECR) with the help of a temporally varying applied magnetic field. The studies have been carried out with the help of Particle-In-Cell simulations. The absorption in the bulk plasma region has been ensured by choosing the value of the initial magnetic field high enough for the laser pulse to propagate inside the plasma region. The magnetic field is then decreased so that the EC resonance takes place inside the plasma. Controlling the temporal profile of the magnetic field, thereafter, the efficiency of energy transfer can be increased. In fact, it is observed that the rate of change of the magnetic field determines the energy transfer process. The other observation shows that by increasing the laser intensity the efficiency of energy transfer process decreases as other processes like harmonic generation etc., are operative which suck energy from the original frequency of the laser pulse.

\section*{Acknowledgements}
The authors would like to acknowledge the OSIRIS Consortium, consisting of UCLA and IST (Lisbon, Portugal), for providing access to the OSIRIS-4.0 framework, which is the work supported by the NSF ACI-1339893. AD would like to acknowledge her J C Bose fellowship grant JCB/2017/000055 and CRG/2022/002782 grant of DST. The authors thank the IIT Delhi HPC facility for computational resources. Rohit Juneja thanks the Council for Scientific and Industrial Research(Grant no. 09/086(1448)/2020-EMR-I) for funding the research.

\bibliographystyle{elsarticle-harv}
\bibliography{Absorption.bib}

\end{document}